\documentclass[conference]{IEEEtran}
\usepackage{cite}
\usepackage{amsmath,amssymb,amsfonts}
\usepackage{algorithm}
\usepackage{graphicx}
\usepackage{textcomp}
\usepackage{color, xcolor}
\usepackage{algpseudocode}
\usepackage{subfigure}
\usepackage{stfloats}
\usepackage{float}

\def\BibTeX{{\rm B\kern-.05em{\sc i\kern-.025em b}\kern-.08em
		T\kern-.1667em\lower.7ex\hbox{E}\kern-.125emX}}
\begin{document}

\title{Seeing is Believing: Detecting Sybil Attack in FANET by Matching Visual and Auditory Domains}

\author{
	\IEEEauthorblockN{Yanpeng~Cui\textsuperscript{*}, Qixun~Zhang\textsuperscript{*}, Zhiyong~Feng\textsuperscript{*}, Xiong~Li\textsuperscript{*}, Zhiqing~Wei{*} and Ping~Zhang\textsuperscript{*}}
	\IEEEauthorblockA{\textsuperscript{*}Key Laboratory of Universal Wireless Communications, Ministry of Education, \\Beijing University of Posts and Telecommunications, Beijing, P.R.China, 100876.\\
		Email: \{cuiyanpeng94, zhangqixun, fengzy, leexiong, weizhiqing, pzhang\}@bupt.edu.cn}}
\maketitle

\begin{abstract}
The flying ad hoc network (FANET) will play a crucial role in the B5G/6G era since it provides wide coverage and on-demand deployment services in a distributed manner. The detection of Sybil attacks is essential to ensure trusted communication in FANET. Nevertheless, the conventional methods only utilize the untrusted information that UAV nodes passively ``heard'' from the ``auditory" domain (AD), resulting in severe communication disruptions and even collision accidents. In this paper, we present a novel VA-matching solution that matches the neighbors observed from both the AD and the ``visual'' domain (VD), which is the first solution that enables UAVs to accurately correlate what they ``see'' from VD and ``hear'' from AD to detect the Sybil attacks. Relative entropy is utilized to describe the similarity of observed characteristics from dual domains. The dynamic weight algorithm is proposed to distinguish neighbors according to the characteristics' popularity. The matching model of neighbors observed from AD and VD is established and solved by the vampire bat optimizer. Experiment results show that the proposed VA-matching solution removes the unreliability of individual characteristics and single domains. It significantly outperforms the conventional RSSI-based method in detecting Sybil attacks. Furthermore, it has strong robustness and achieves high precision and recall rates.
\end{abstract}


\section{Introduction}
The Unmanned Aerial Vehicle (UAV) network has been envisioned as a promising solution for numerous applications since it provides wide coverage and on-demand deployment services for the B5G/6G mobile communication. The high deployment cost and poor efficiency of the centralized UAV networks could be eliminated via the distributed and collaborative self-organized one, the flying ad-hoc network (FANET) thus emerges. However, UAV nodes in FANET are continuously exposed to attacks that might cause severe accidents \cite{UAV_Safety}. For instance, in Kamkar’s Skyjack project, a malicious node easily intruded on UAVs with disguising identities, which forcibly disconnected them from the legitimate transmitter and arbitrarily fed commands to all possessed zombie UAVs \cite{Skyjack}. Besides, with the flight control system being hacked, several UAVs rained down on civilians during an anniversary celebration of a supermarket in China’s Zhengzhou city last year \cite{Zhengzhou}. These issues might have been avoided if UAVs could accurately identify the disguised nodes. The requirement for trusted communication thus naturally arises in FANET.

Utilizing the shared medium and the broadcast nature of wireless communication, malicious UAVs usually generate several virtual and fake identities to intrude and disturb a considerable portion of FANET, which is known as the Sybil attack \cite{Sybil_Attack}. It provides chances for disruptions in resource allocation, vote mechanisms, time synchronization, and routing decisions \cite{UAV_Sybil_Attack}. Based on symmetric keys protocols, blockchain techniques have been utilized for Sybil detection. Whereas the time consumption of the transaction execution can not be overlooked in the case of a large blockchain-based network. For instance, the blockchain-assisted solution \cite{Key2} consumes 2000ms in a FANET with 50 UAV nodes. It is intolerable since the topology of FANET changes rapidly. Given this, a fast and secure group key establishment protocol is proposed in \cite{Group_Key}, and another temporal credential-based anonymous lightweight user authentication mechanism for UAV networks is presented in \cite{TCALAS}, which provides lower costs in both computation and communication. Whereas the compromise of the authentication server would result in the exposure of keys when conducting such centralized schemes.

To address the above-mentioned issues, some lightweight and distributed Sybil detection methods that are built upon other principles are well underway, including radio resource tests-, relative distance-, neighbor information-, mobility- and energy-based techniques. However, most of them are dedicated to mobile ad hoc networks with low mobility and are unsuitable for UAV networks for the following reasons. The radio resource test \cite{Channel_Test} does not work if malicious UAVs utilize multiple radio devices simultaneously. The received signal strength indicator (RSSI) scheme \cite{RSSI_Based} fails when the malicious nodes shift their transmission powers. The neighbor-based solution \cite{Neighbor_Based} loses its efficacy in a dense network since two adjacent legitimate UAVs will be misjudged owing to the same neighbor list. The mobility-based technique \cite{Mobility_Based} may fail since a set of legitimate UAVs flying in formation will be erroneously identified as Sybil nodes. The energy-based method \cite{Energy_Based} may not work since a malicious UAV may mutely monitor the behavior of legitimate nodes and keep a record of them, and then impersonate their identity and level of energy. Despite years of intensive research, the limitation of the aforementioned solutions stems from the following fact: UAVs only utilize the information they passively obtain, which has poor reliability.

In essence, the above issues occur since the neighbor's information is only passively ``heard" from the radio domain, i.e., UAVs are communicating with eyes closed. Following the spirit of ``seeing is believing", they should have the capability to open their eyes when communicating with others \cite{Dual_Identity} \cite{UAV_ISAC2}. Compared with the ``auditory domain (AD)'', the physical characteristics observed in the ``visual domain (VD)'' are difficult to be disguised, which enhances the detection reliability of the Sybil attacks. For example, one malicious UAV can effortlessly disguise multiple Sybil identities in AD, but it is difficult to disguise in VD since such identities do not exist in the real world \cite{Magazine}. To this end, UAVs should be capable of additionally observing the neighbor's physical characteristics in VD to realize trusted communication. 

The exciting part is that VD-based sensing is easy to accomplish since state-of-the-art techniques have been utilized on UAVs to enable active sensing ability. The most typical ones include Lidar, high-resolution cameras, circular scanning millimeter-wave (CSM) Radar, and laser range finder (LRF) \cite{UAV_Sensors}. Nevertheless, the information obtained in AD and VD is still separated thus far. In addition, it is not trivial to match what the UAVs hear and see due to the \textbf{Similarity} issue. That is, the physical characteristics of neighbors tend not to differ much sometime or somewhere. For instance, the difference in relative velocity is more distinctive if UAVs fly freely, whereas paltry when they are in formation. \textbf{Uncertainty} is another issue since most physical characteristics are observed with specific deviations. These issues may induce difficulty in distinguishing or even mismatches when associating observations from both domains.

The above issues and challenges motivate us to fully utilize the information from dual domains to detect the Sybil attacks. This paper presents a lightweight Sybil detection solution by matching the neighbor information obtained from VD and AD, which is called VA-matching. To the best of our knowledge, this is the first solution that enables UAVs to accurately correlate what they ``see'' and ``hear'' to improve the detection accuracy of the Sybil attacks. To address the \textbf{Uncertainty} issue, we utilize relative entropy (RE) to describe the similarity of the densities of physical characteristics. Aiming at solving the \textbf{Similarity} issue, we propose a dynamic weight algorithm to distinguish various physical characteristics and calculate the matching cost via their popularity. The matching model is established by a bipartite graph and solved by the vampire bat optimizer (VBO). The main benefits of VA-matching are two folds. i) It removes the low detection accuracy of Sybil attacks induced by inaccurate and untrusted information obtained from a single domain, especially from AD. Consequently, it outperforms the AD-only method in detection precision. ii) The rates of precision and recall are identical in VA-matching. Therefore, one can achieve the best performance without the other being compromised, which is challenging for conventional solutions.

\section{System Model}
\subsection{Motivation Scenario}
\begin{figure}
	\centering
	\includegraphics[width=1\linewidth]{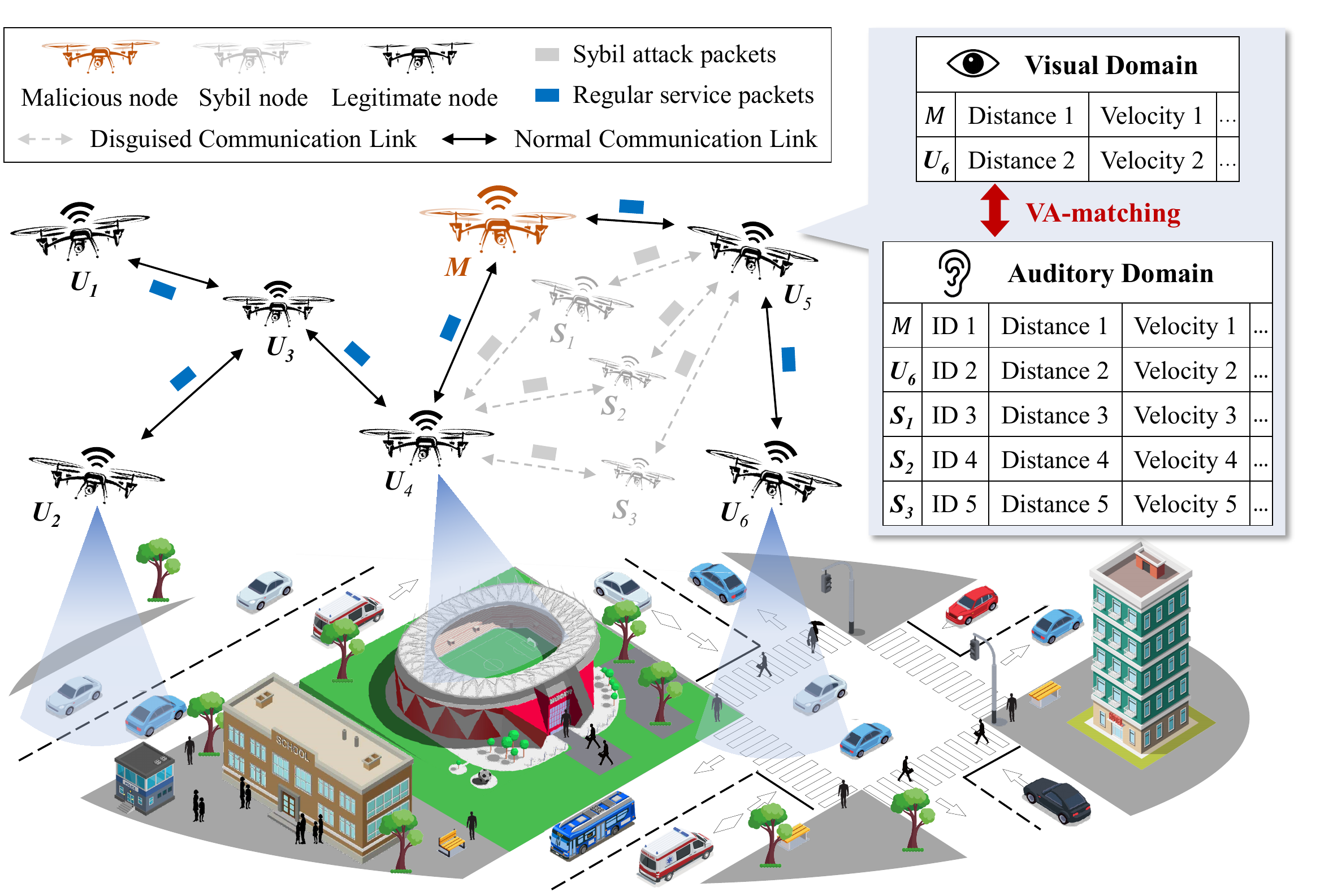}
	\caption{The typical scenario for Sybil attacks in FANET.}
	\label{Scenario}
\end{figure}
As present in \textbf{Fig. \ref{Scenario}}, a FANET consisting of multiple UAVs is deployed to provide aerial monitoring and high-capacity wireless communication for the smart city. With the region of interest appearing randomly, UAV nodes move freely, i.e., adjust the destination, moving speed, and heading direction independently. To realize the communication functionality, each UAV is equipped with an omnidirectional antenna above the rotor. In addition to passively receiving messages from AD, high-resolution cameras, CSM Radar, and LRF are also equipped to enable active sensing ability from VD. Similar to most UAV-related studies \cite{Z. Xiao} \cite{TARRAQ}, we assume that the link between UAVs is mainly dominated by the line of sight (LoS) channel\footnote{Our solution can be easily extended to the Non-LoS (NLoS) and multi-antennas scenarios. Given the page limit, it won't be reiterated here. }. There are three categories of nodes involved in the motivation scenario, namely legitimate UAVs, malicious UAVs and their Sybil identities. A malicious UAV attempts to attack its one-hop neighbors by utilizing multiple disguised nodes, namely the Sybil identities. The legitimate UAVs intend to detect and find the Sybil identities.

\subsection{Network Model}

Assuming there are $N$ UAVs moving in a cube region with a length of $L$, a width of $W$, and a height of $H$. All UAVs are aware of their own location and velocity via a global navigation satellite system and inertia surveying system within tolerable error. Consequently, each UAV has the capability to measure a location vector $\textbf{\textit{p}}_n=[p_n(1),p_n(2),p_n(3)]^T$ and a velocity vector $\textbf{\textit{v}}_n=[v_n(1),v_n(2),v_n(3)]^T$ of itself. The mobility information will be embedded in beacons and periodically exchanged, and also exploited to perform VA-matching. Recall that the LoS links dominate the UAV communication links, the path loss model between the $i$th UAV $U_i$ and the $j$th UAV $U_j$ is assumed to follow $d(i,j)^{-\alpha}$, where $\alpha$ is the mean value of path loss exponent and $d(i,j)$ is the Euclidean distance between $U_i$ and $U_j$.

The Signal to Interference plus Noise Ratio (SINR) from $U_i$ to $U_j$ is given by $\gamma(i,j)=\frac{P(i)h(i,j)d(i,j)^{-\alpha}}{N_0+N_I},$
where $P(i)$ is the transmission power of $U_i$, and $N_0$ is the additive white Gaussian noise. The power gain of small-scale fading channel $h(i,j)$ is assumed to be exponentially distributed with a unit mean. According to \cite{Srinivasa}, the interference is given by $N_I=\!\sum\nolimits_{k\neq i,j}\!P_{k}h(k,j)d(k,j)^{-\alpha}\!=\!\frac{3N_e(D_m^{3-\alpha}-D_s^{3-\alpha})}{2D_m^3(3-\alpha)}$,
where $N_e=4\pi D_m^3N/(3L\times W\times H)$ is the equivalent number of UAVs in a sphere with radius $D_m=\sqrt{L^2+W^2+H^2}$, which is the maximum distance between any two UAVs. $D_s$ is the safe distance of UAVs to avoid collisions. Therefore, the outage probability ${\rm Prob}\left(\gamma(i,j)\ge \gamma_{th}\right)$ is given by ${\rm P_o}(i,j)={\rm Prob}\left\{ {h(i,j)} \ge\gamma_{th}d(i,j)^{\alpha}(N_0+N_I)/P(i)\right\} = \exp\left(-\gamma_{th}d(i,j)^\alpha(N_0+N_I)/P(i)\right)$,
where $\gamma_{th}$ is the SINR threshold. The condition ${\rm P_o}\ge{\rm P}_{th}$ should be satisfied to ensure the communication demands, where ${\rm P}_{th}$ is the constraint on SINR probability. Therefore, the effective transmission distance can be calculated by $D_r(i,j) = \left(-\frac{P(i)\ln \left( {\rm P}_{th} \right)}{\gamma_{th}(N_0 + N_I)} \right)^{\frac{1}{\alpha}}$.

\subsection{Attack Model}
We assume that there are a certain number of malicious nodes in the network, accounting for $P_m$ of the total number of UAV nodes. Each malicious node can create $N_s$ Sybil identities, and make direct, simultaneous, and fabricated attacks with the following implications. 
\textit{1) Direct attack:} a malicious UAV along with its Sybil identities broadcast beacons to show its presence, and a legitimate node receives the beacon and then responds via a one-hop link. Consequently, the Sybil nodes directly attack the legitimate node by pretending its one-hop neighbors. 
\textit{2) Simultaneous attack:} At the $n$th epoch, the malicious UAV introduces all Sybil identities at once. 
\textit{3) Fabricated attack:} malicious UAV fabricates some fake identities that do not exist previously, rather than stealing the existing identities of legitimate nodes. 
\textbf{Fig. \ref{Scenario}} intuitively presents an example, where a malicious UAV $M$ disguises three Sybil nodes $S_1$, $S_2$, and $S_3$, and forces the legitimate nodes $U_4$ and $U_5$ to mistake them as one-hop neighbors.

\section{Detecting Sybil Attacks by VA-matching}
\begin{figure}
	\centering
	\includegraphics[width=0.95\linewidth]{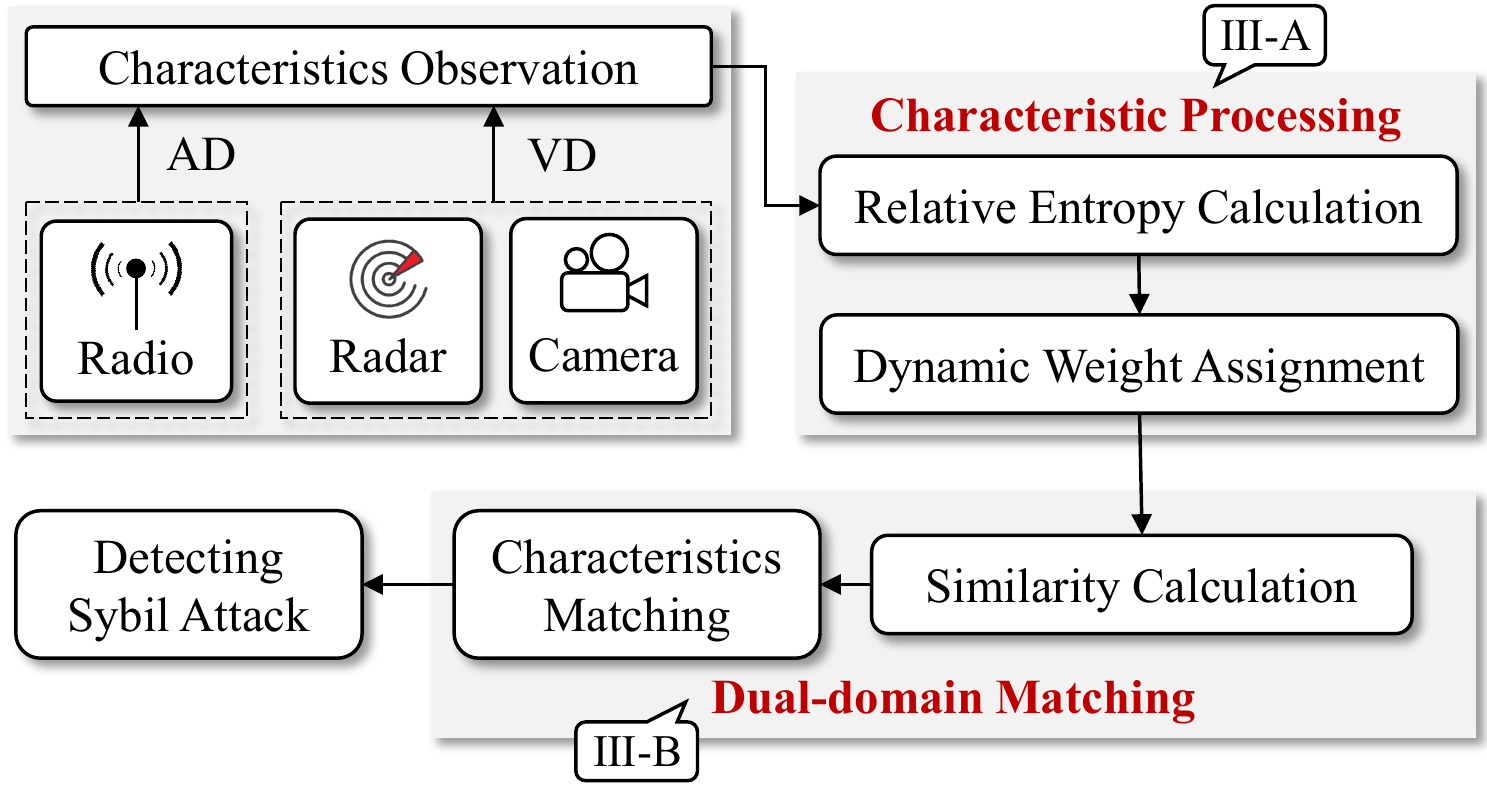}
	\caption{The framework of VA-matching. The critical steps are shown in red and presented in Sec. \textbf{\uppercase\expandafter{\romannumeral3}-A} and \textbf{\uppercase\expandafter{\romannumeral3}-B}.}
	\label{Framework}
\end{figure}

As shown in \textbf{Fig. \ref{Framework}}, the proposed VA-matching solution is started by letting each UAV observe the physical characteristics of neighbors from dual domains. Recall that AD only refers to wireless communication while VD contains all the active sensing methods, e.g., Radar and cameras, etc. At the $n$th epoch, a UAV measures neighbors’ physical characteristics $\textbf{VF}^{[n]}=\{\textbf{\textit{vf}}_i^{[n]},i=1,...,K_v\}$ and $\textbf{AF}^{[n]}=\{\textbf{\textit{af}}_j^{[n]},j=1,...,K_a\}$, where $K_v$ and $K_a$ denote the number of neighbors observed in VD and AD, respectively. $\textbf{\textit{vf}}_{i,k}^{[n]}$ and $\textbf{\textit{af}}_{j,k}^{[n]}$, $k=1,...,K_f$ denote the $k$th physical characteristic of the $i$th neighbor in VD and the $j$th neighbor in AD, respectively. $K_f$ is the total number of observable characteristics. Subsequently, each UAV processes the obtained characteristics and matches the neighbors in dual domains, and finally detects the Sybil attacks. The specific scheme is as follows.
\subsection{Characteristic Processing}

The physical characteristics obtained from AD and VD will be extracted to get further process. Their reliability varies greatly since they generally deviated from the true value with Gaussian error. Furthermore, the reliability depends on their popularity. For instance, the relative distance is perfect to distinguish neighbors in different locations while the relative velocity is not a good characteristic if neighbors are moving in formation. Therefore, the similarity of any two characteristics is calculated by RE to solve the \textbf{Uncertainty} issue. They are also assigned with dynamic weights according to their popularity to address the \textbf{Similarity} issue. The characteristic processing includes the following two folds.

\subsubsection{Relative Entropy Calculation}

The RE of two probability density functions $p(x)$ and $q(x)$ is defined as $R(p||q)=\int p(x){\rm ln}\left(p(x)/q(x)\right)$.
It measures the ineffectiveness when the true distribution is $p$ and the assumed distribution is $q$.
Therefore, the similarity of the two characteristics can be determined by calculating the pair of probability density functions (PDF). Note that the observation results can be written in Gaussian forms, which can be seen from the experimental results in Sec. \textbf{\uppercase\expandafter{\romannumeral4}-A}, so the RE of two characteristics $\textbf{\textit{f}}_a$ and $\textbf{\textit{f}}_b$ is given by
\begin{equation}\label{RE}
	\begin{aligned}
		R(\textbf{\textit{f}}_a, \textbf{\textit{f}}_b)=&
		\frac{{\rm ln}\left(\sigma_{\rm b}/\sigma_{\rm a}\right)}{2\sigma_{\rm b}^2\sigma_{\rm a}^2\sqrt{2\pi}\sigma_{\rm b}}
		\!\int{\rm exp}\left(-\frac{(x-\textbf{\textit{f}}_a)^2}{2\sigma_{\rm a}^2}\right)\times\\
		&\left(\sigma_{\rm a}^2(x-\textbf{\textit{f}}_b)^2-
		\sigma_{\rm b}^2(x-\textbf{\textit{f}}_a)^2\right)dx,
	\end{aligned}
\end{equation}
where $\sigma_{\rm a}^2$ and $\sigma_{\rm a}^2$ are the variance of $\textbf{\textit{f}}_a$ and $\textbf{\textit{f}}_b$, respectively. Due to the asymmetric problem of RE, namely $R(p,q)\neq R(q,p)$, the $R(\textbf{\textit{vf}}_{i,k}, \textbf{\textit{af}}_{j,k})$ can not be directly utilized in the subsequent matching procedure. The similarity of $\textbf{\textit{vf}}_{i,k}$ and $\textbf{\textit{af}}_{j,k}$ should be the same as that of $\textbf{\textit{af}}_{j,k}$ and $\textbf{\textit{vf}}_{i,k}$. We consider a Jensen–Shannon divergence $D(\textbf{\textit{f}}_a, \textbf{\textit{f}}_b)=R(\textbf{\textit{f}}_a, \textbf{M})/2 + R(\textbf{\textit{f}}_b,\textbf{M})/2$
to measure the similarity between $\textbf{\textit{f}}_a$ and $\textbf{\textit{f}}_b$, where $\textbf{M} = (\textbf{\textit{f}}_a+\textbf{\textit{f}}_b)/2$.

\subsubsection{Dynamic Weight Assignment}
By exploiting the RE, the similarity of two physical characteristics can be calculated. When they come from the same domain, $D(\textbf{\textit{f}}_a, \textbf{\textit{f}}_b)$ describes the similarity of physical characteristics of two UAVs. When they come from various domains, $D(\textbf{\textit{f}}_a, \textbf{\textit{f}}_b)$ describes how likely two physical characteristics belong to the same UAV. Assume that $\textbf{AF}_k=\{\textbf{\textit{af}}_{1,k},...,\textbf{\textit{af}}_{j,k},...,\textbf{\textit{af}}_{K_a,k}\}$ denotes the measurement of the $k$th physical characteristics of $K_a$ neighbors in AD. The weight of $k$th characteristic is assigned as $w_k=\frac{1}{K_a}\sum\nolimits_{j=1}^{K_a}p_{j,k}$,
where $p_{j,k}$ denotes the distinguishability of $\textbf{\textit{af}}_{j,k}$, i.e., the probability that $\textbf{\textit{af}}_{j,k}$ is different from other $K_a-1$ characteristics in $\textbf{AF}_k$. The smaller it is, the more popular the $k$th physical characteristic is, and its contribution is expected to be smaller when calculating the similarity. Therefore, $p_{j,k}$ is defined as $p_{j,k}=\sum\nolimits_{p\neq j}D\left(\textbf{\textit{af}}_{j,k}, \textbf{\textit{af}}_{p,k}\right)\times\prod\nolimits_{q\neq j,p}(1-D(\textbf{\textit{af}}_{j,k}, \textbf{\textit{af}}_{q,k}))$, 
where $D\left(\textbf{\textit{af}}_{j,k}, \textbf{\textit{af}}_{p,k}\right)\prod\nolimits_{q\neq j,p}(1-D(\textbf{\textit{af}}_{j,k}, \textbf{\textit{af}}_{q,k}))$ denotes the probability that $\textbf{\textit{af}}_{j,k}$ is the same as $\textbf{\textit{af}}_{p,k}$ but different from the other measurements in $\textbf{AF}_k$. In this way, the characteristic with a high distinguishability will be assigned with a large weight, and the \textbf{Similarity} issue could be solved commendably.
\subsection{Dual-domain Matching}
\subsubsection{Similarity Calculation}
Based on the characteristic processing procedure, the similarity of $\textit{\textbf{vf}}_i^{[n]}$ and $\textit{\textbf{af}}_j^{[n]}$ can be calculated. We defined it as the harmonic mean of individuals $s_{i,j}^{[n]}=\big\{\sum\nolimits_{k=1}^{K_f}\tilde{w}_k^{[n]}D^{-1}\big(\textbf{\textit{vf}}_{i,k}^{[n]}, \textbf{\textit{af}}_{j,k}^{[n]}\big)\big\}^{-1}$, 
where $\tilde{w}_k^{[n]}$ is the normalized weight. In this way, the physical characteristics will be dynamically distinguished according to their popularity. It assigns small weights to characteristics that are hard to distinguish the neighbors. For instance, if all neighbors are moving in formation, we'll have $p_{i,k}\rightarrow0$ for the characteristic of relative velocities since they hardly have any difference, and thus the corresponding weight will tend to zero. Consequently, the relative velocity will almost have no contribution when measuring the similarity of neighbors. In addition, if all the relative locations are distinguishable enough, then the weight for distance will be relatively large and it thus yields a higher weight and plays a more important role. In addition, another crucial property of harmonic mean is that when $\tilde{w}_k^{[n]}$ is given, $s_{i,j}^{[n]}$ will tend to zero if any $D(\textbf{\textit{vf}}_{i,k}^{[n]}, \textbf{\textit{af}}_{j,k}^{[n]})$ tends to zero. In other words, if two neighbors have high dissimilarities in most characteristics, the similarity will be small despite the large weights of other characteristics. This property can effectively eliminate the impact of outliers.

\subsubsection{Characteristics Matching}
\begin{figure}
	\centering
	\includegraphics[width=0.9\linewidth]{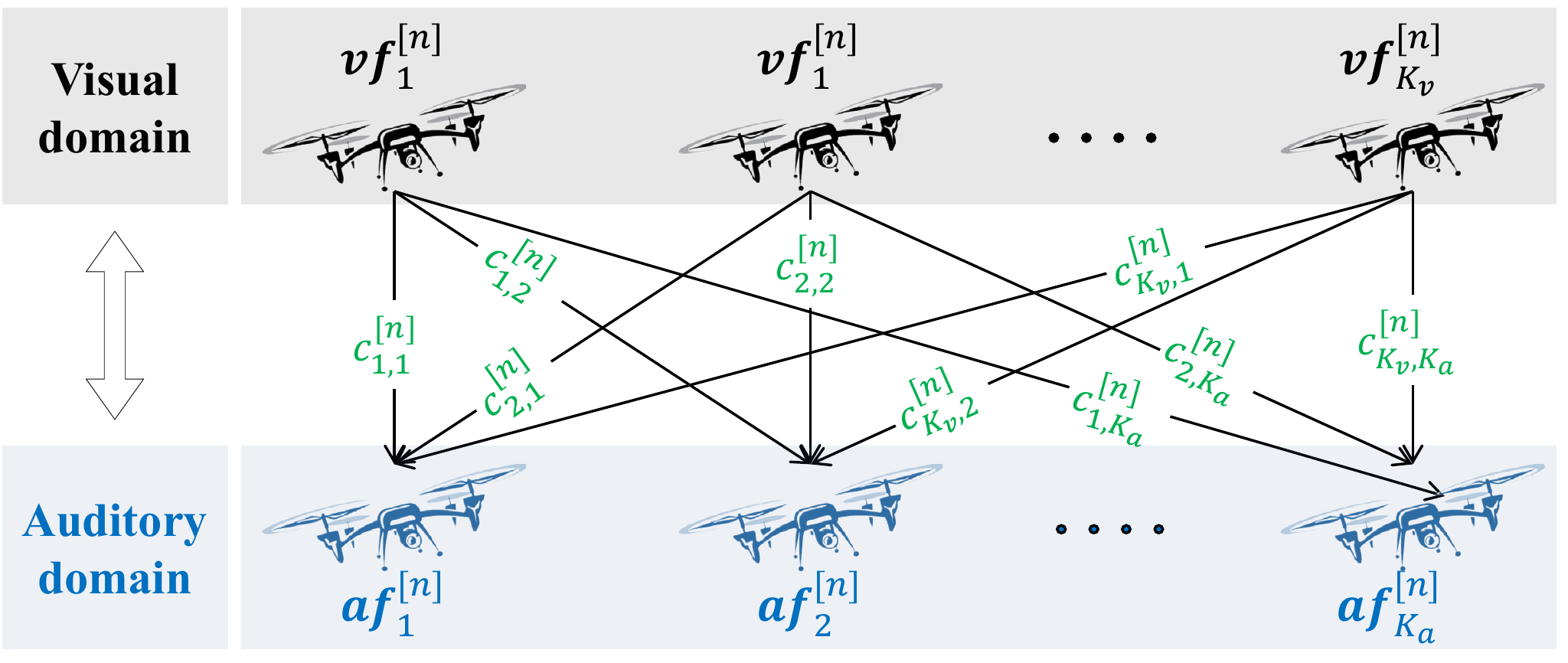}
	\caption{Bipartite graph model for matching neighbors in VD and AD.}
	\label{Matching_Model}
\end{figure}

The physical characteristics obtained from AD and VD will be exploited to establish a bipartite graph matching model shown in \textbf{Fig. \ref{Matching_Model}}. The edge's weight is defined as the matching cost $c_{i,j}^{[n]}=1/s_{i,j}^{[n]}$, namely the reciprocal of similarity between $\textit{\textbf{vf}}_i^{[n]}$ and $\textit{\textbf{af}}_j^{[n]}$. To minimize the global cost and balance the individual cost, the optimization target of dual-domain matching is given by $\min(f_1+f_2)$, where $f_1=\sum\nolimits_{i=1}^{K_v}\sum\nolimits_{j=1}^{K_a}a_{i,j}^{[n]}c_{i,j}^{[n]}$ and $f_2=\sum\nolimits_{i=1}^{K_v}\sum\nolimits_{j=1}^{K_a}\big(a_{i,j}^{[n]}c_{i,j}^{[n]}-f_1\big)^2$
are the sub-problems about the global and individual costs, respectively. The constraints are defined as $\sum\nolimits_{i=1}^{K_v}a_{i,j}^{[n]}=1, j=1,...,K_a$ and $\sum\nolimits_{j=1}^{K_a}a_{i,j}^{[n]}=1, i=1,...,K_v$,
which indicates that any physical characteristic obtained from AD can be only matched with the other one in VD and vice versa. $\textbf{\textit{a}}^{[n]}$ is the assignment matrix, where $a_{i,j}^{[n]}=1$ if $\textbf{\textit{vf}}_i^{[n]}$ is assigned to $\textbf{\textit{af}}_j^{[n]}$, otherwise $a_{i,j}^{[n]}=0$. Note that there are more neighbors observed in AD than in VD since the Sybil node only exists in the former, thus the existing assignment algorithm cannot be used directly since the cost matrix is generally not a square one. We thus expand it by adding $K_a-K_v$ lines of infinite elements below $\textbf{\textit{C}}^{[n]}$.

The above-mentioned problem of matching neighbors in dual domains can be solved by the VBO, which is an effective optimizer we recently proposed. The specific steps of VBO will not be introduced here due to page limitations, and readers can refer to our recent work \cite{Vampire Bat Optimizer} for more details.

\subsection{Detecting Sybil Attacks}
Our aim is to detect the Sybil identity created by malicious UAVs. Here, we will explain how the proposed VA-matching completes this task. Since the legitimate UAV is within the one-hop communication range of the malicious UAV, it therefore mistakenly believes that the disguised Sybil nodes are the actual neighbor around. Therefore, the number of the neighbor list obtained in AD and VD is unequal. By comparing the number of neighbors in AD and VD, some additional nodes will be found in AD but we still can not tell which are Sybil nodes. That's the reason why the VA-mapping solution is designed to find the neighbors in AD that are most similar to those in VD. Finally, after matching the characteristics of neighbors in dual domains, the remaining unmatched neighbors in AD (if any) will be determined as Sybil nodes.

\subsection{Computational Complexity}

Here we analyze the computational complexity of the proposed VA-matching. Note that the procedures before VBO mainly include the dynamic calculation of the similarity of dual domain neighbors, which depends on the number of neighbors and the observed characteristics. Specifically, there will be $(K_a-1)^2$ calculations before determining the distinguishability of each physical characteristic of neighbors, namely the complexity is $O(K_a^2)$. Based on this, the similarity calculation of the features observed in AD and VD consumes another $K_f$ iterations. Therefore, the total complexity is $O(K_fK_a^2)$. In addition, the VBO has the order of complexity $O(K_a^2{\rm log}(K_aC))$ to equalize and minimize the local cost, where $C$ is the maximum absolute value of competition. Considering that the $K_f$ is sufficiently smaller than ${\rm log}(K_aC)$ in the large-scale FANET, the computational complexity of the proposed VA-matching solution can be thus regarded as that of the VBO.

\section{Experiment and Simulation}
\subsection{Real-World Experiment}
\begin{figure}
	\centering
	\subfigure[DJI M300 RTK]{\includegraphics[width=0.45\columnwidth]{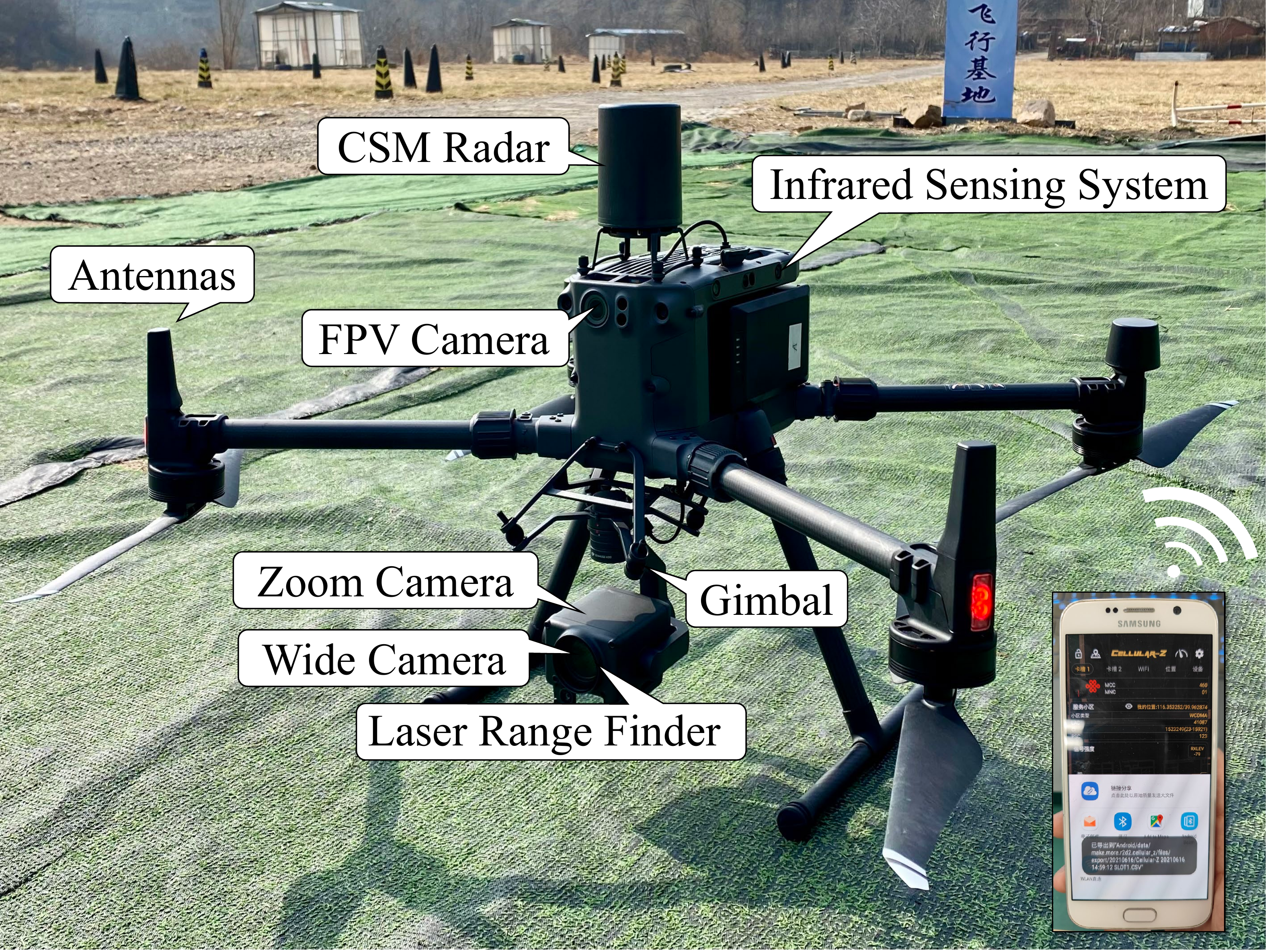}}\quad
	\subfigure[Detection result of ZF-F1200]{\includegraphics[width=0.45\columnwidth]{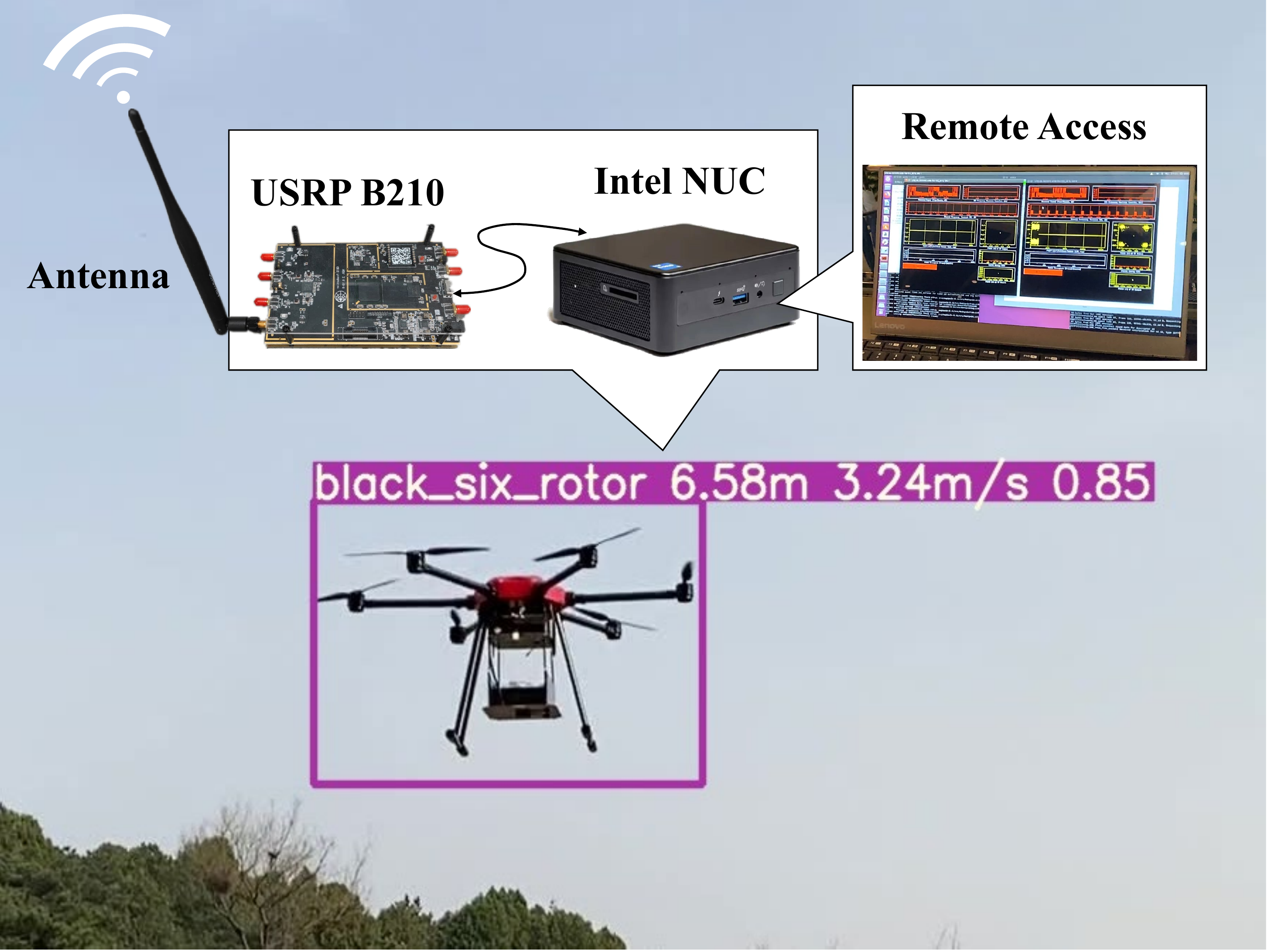}}
	\caption{UAV-based testbed for real-world experiments.} \label{Real_Experiments}
\end{figure}

\begin{figure}
	\centering
	\includegraphics[width=1\linewidth]{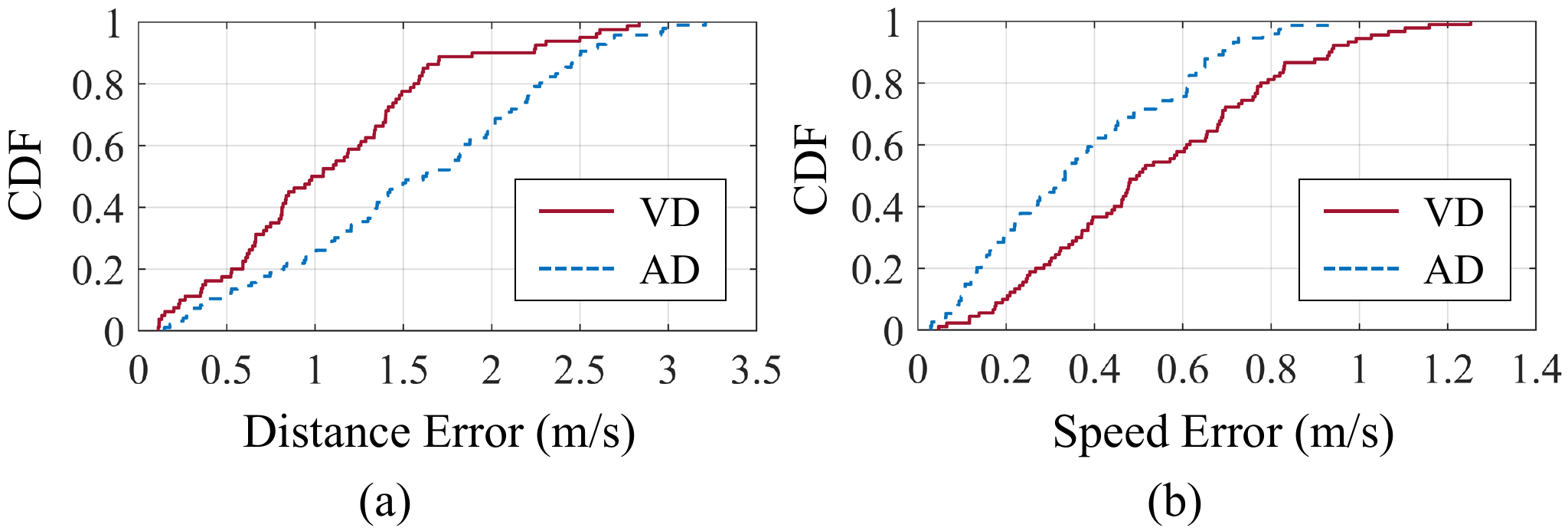}
	\caption{Real-world experiment results: the distribution of observation errors of (a) distance and (b) speed in AD and VD.}
	\label{CDF}
\end{figure}
The real-world experiment aims to determine the measurement error of UAV's characteristics in both AD and VD. As shown in \textbf{Fig. \ref{Real_Experiments}(a)}, we deploy CSM Radar, Zenmuse H20 gimbal \& camera system, and infrared sensing system on DJI Matrice 300 (M300) RTK. To build a VD-based measurement approach, we applied YOLO version 5 on M300 RTK to detect the relative distance and velocity of ZF-F1200, a six-rotor UAV shown in \textbf{Fig. \ref{Real_Experiments}(b)}. Instead of exchanging beacons between them, we calculate the relative velocity and distance based on the exported data of positioning and speed sensors of M300 and ZF-F1200, and regard them as characteristics obtained in AD\footnote{Even if the beacons are exchanged, the embedded states are derived from sensors. So we omit the beacon exchange to simplify the experiment.}. It can be seen from \textbf{Fig. \ref{CDF}} that AD has a lower measurement error of speed but a higher one of distance. We also find that their errors follow Gaussian distributions. Specifically, when using the Gaussian distribution based on the mean and variance in \textbf{Table \ref{Table}}, the experimental results are fitted with an average probability error below 0.0005. This means that they follow the normal distribution, which provides the basis of experimental data for the formula (\ref{RE}).

\begin{figure}
	\centering
	\includegraphics[width=0.85\linewidth]{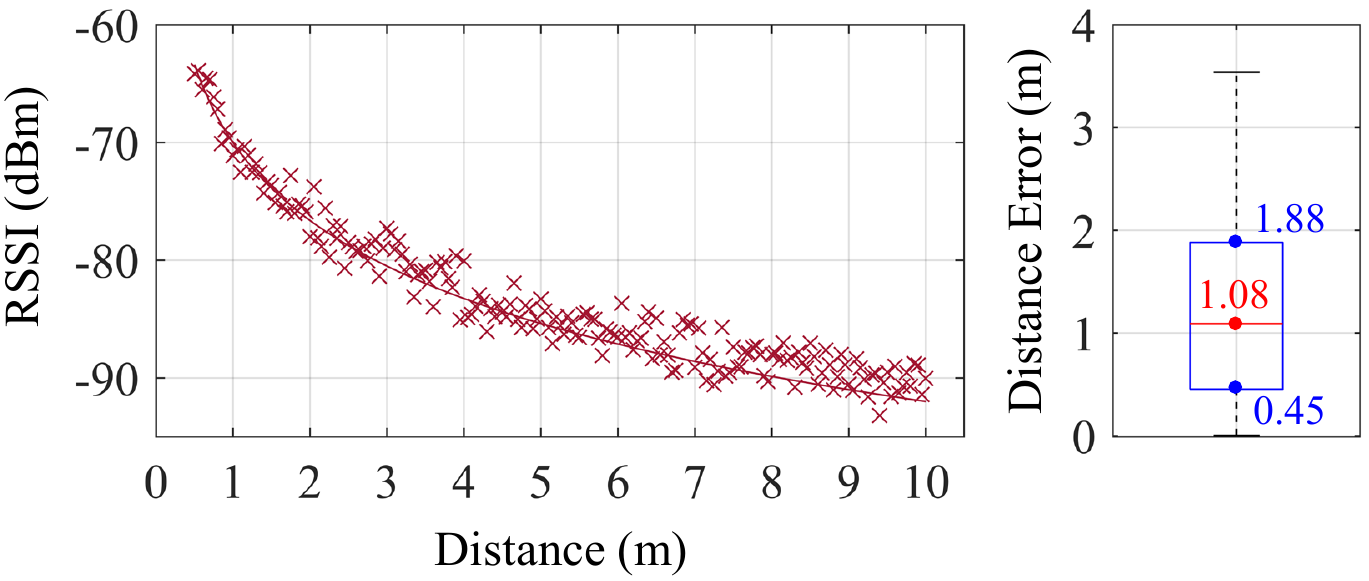}
	\caption{Real-world experiment results: (a) the relationship of RSSI and distance; (b) ranging error.}
	\label{RSSI}
\end{figure}

To build an RSSI-based Sybil detection as a comparison scheme for the subsequent simulation, we establish an error model for the RSSI-based ranging method. An Ettus USRP B210 connected to a mini-computer with OpenAirInterface (OAI) platform is developed in the pod of ZF-F1200. A Samsung S6 mobile phone is fixed in M300's pod to receive the signal transmitted by the OAI platform. The RSSI is measured and recorded via the Cellular-Z application in Samsung S6. The ranging error of RSSI is shown in \textbf{Fig. \ref{RSSI}}. Within the range of 10m, the median value is 1.08m, and the upper and lower quartiles are 1.88m and 0.45m, respectively. The average probability loss is the smallest (remains below 0.0025) when fitting the error by a normal distribution with a mean value of 1.26m and a variance of 0.86m.

\begin{table}
	\centering
	\caption{The parameters of Gaussian distribution that fits the results of the real-world experiment.}
	\includegraphics[width=0.9\linewidth]{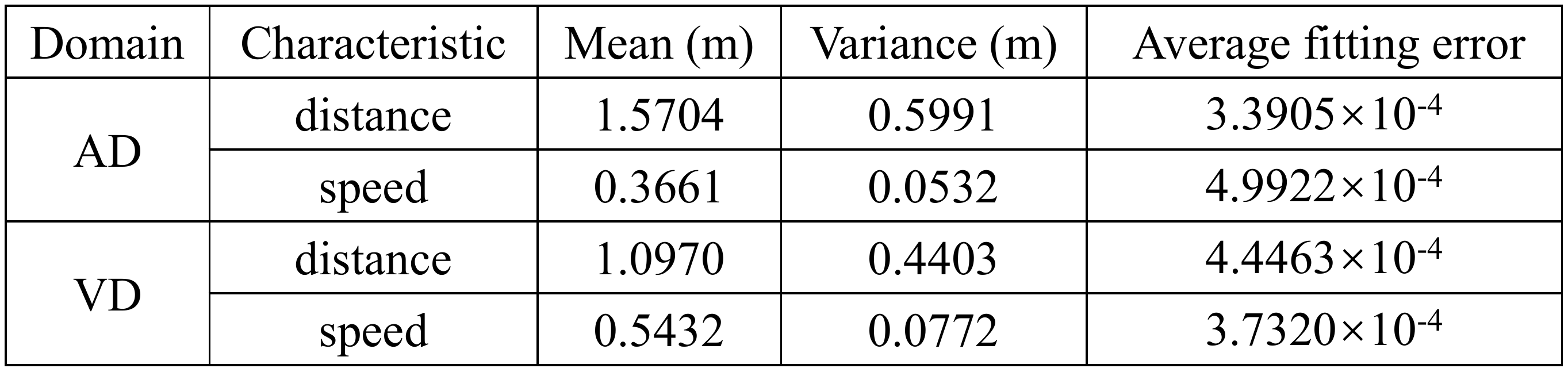}
	\label{Table}
\end{table}

\subsection{Simulation Results}
In the large-scale simulations, we set $\alpha = 2$, ${\rm P}_{th}=0.8$, $D_s=5$m for the communication model. The transmit power of each UAV node is 30 dBm. UAVs are moving with
the random waypoint model in a region of 600m$\times$600m$\times$300m and 400m$\times$400m$\times$200m. The number of UAV nodes is between 20 and 150, and their lowest speed is 5m/s while the maximum speed varies from 10m/s to 20m/s. The SINR threshold varies from -10dB to -4dB. The proportion of malicious nodes $P_m$ is set as 0.1 and 0.2, and each of them could generate 10 Sybil identities. The total simulation time is 300s and the results are collected every 2s. The error model of RSSI-based ranging as well as the observation in AD and VD are from the results of the real-world experiment. The competing rate is set as 1 and $\epsilon=0.02$ in the VBO. Simulations are run 20 times and the results with 90\% confidence are analyzed as follows.

\subsubsection{Matching Accuracy}

\begin{figure}
	\vspace{0.03in}
	\centering
	{\includegraphics[width=0.7\columnwidth]{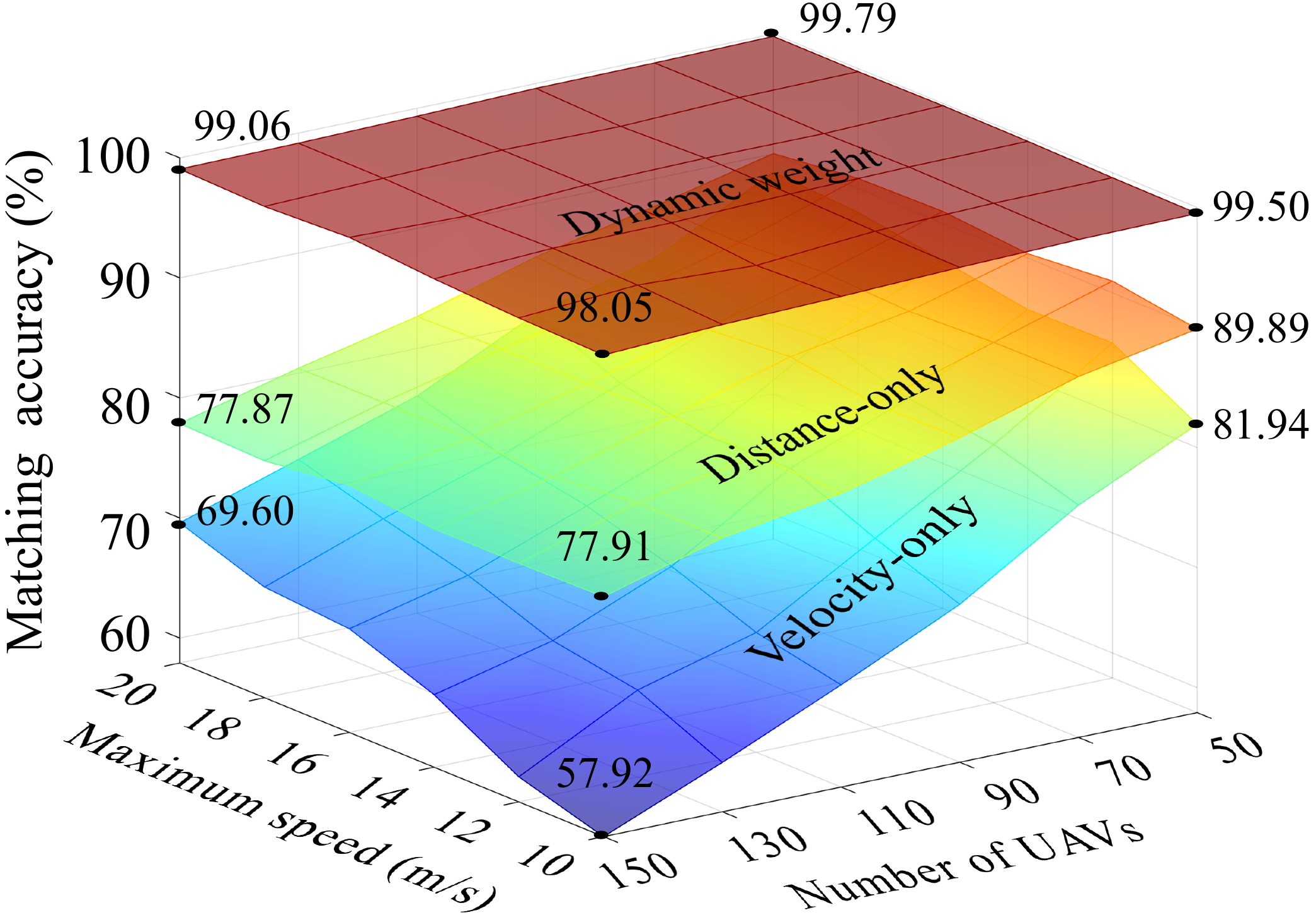}}
	\caption{Matching accuracy of various solutions under different scenarios.}\label{Matching_Rate}
\end{figure}

Since the proposed VA-matching is working based on matching theory, we first evaluate the matching accuracy, which is defined as the proportion of correctly matched neighbors in the dual domain. As shown in \textbf{Fig. \ref{Matching_Rate}}, the distance-only and velocity-only schemes, whose cost matrix is only generated based on distance or speed, are used for comparison. They can be regarded as the case where our scheme only utilizes one characteristic, i.e., weights for others are 0. The proposed dynamic weight outperforms the distance-only and velocity-only methods by 15.12\% and 25.43\% on average, respectively. As expected, our proposal has the best performance, namely over 98.05\% matching accuracy under all maximum speeds and network densities, since it distinguishes the popularity of characteristics and assigns dynamic weights. In contrast, the distance-only method only uses the relative distances of neighbors regardless of their velocities. As a result, it is worse than our VA-matching and is sensitive to the total number of UAVs. The denser the network, the more neighbors each UAV will have, resulting in a higher possibility of similar relative distances and therefore a worse matching accuracy. The velocity-only method, as the name implies, only utilizes the velocity feature to generate the cost matrix. It is sensitive to the network density for the same reason as the distance-only one. Furthermore, because the minimum speed is limited to 5m/s, a lower upper-speed limit means that there is a greater probability of neighbors with similar speeds. Consequently, in a network with 150 UAV nodes, the matching accuracy of the velocity-only method is lower than 60\% when the upper-speed limit is 10m/s. 

\subsubsection{Performance of Detecting Sybil Attacks}
\begin{figure}
	\centering
	\includegraphics[width=0.8\linewidth]{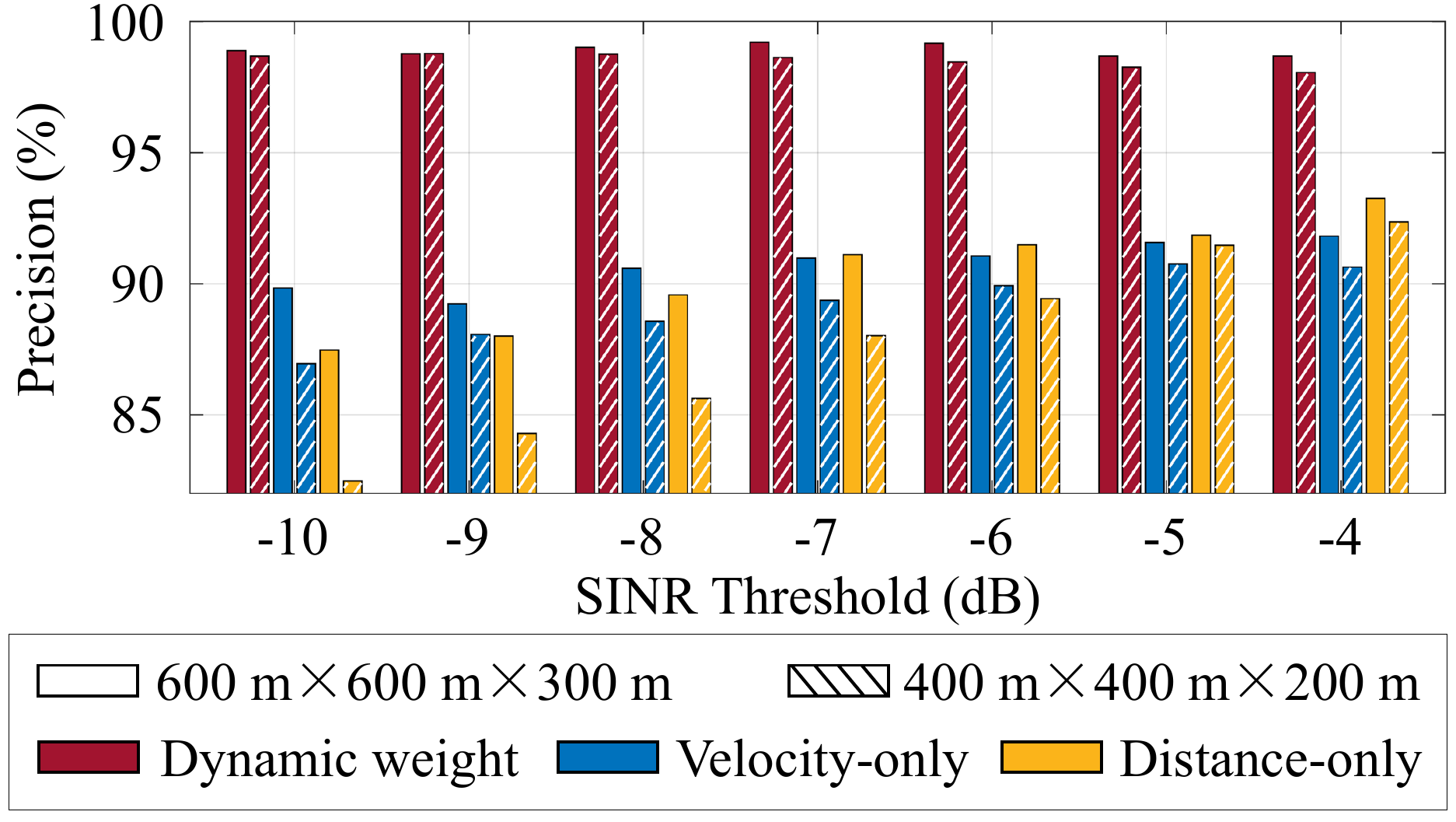}
	\caption{Detection precision of various solutions under different scenarios.}
	\label{Precision_VS_SINR}
\end{figure}

To a certain extent, the matching accuracy will affect the detection precision of Sybil attacks since the unmatched neighbors in AD will be regarded as Sybil nodes. As shown in \textbf{Fig. \ref{Precision_VS_SINR}}, we evaluate the precision of Sybil attacks under various simulation environments. The detection precision of the Sybil attacks is defined as P=TP/(TP+FP), where the true positive (TP) means the number of the actual Sybil nodes that are successfully detected, while the false positive (FP) denotes the number of nodes that are misjudged as the Sybil nodes. All schemes have lower precision in small-size networks since the number of neighbors will increase, leading to the possibility of similar characteristics and making it more difficult to distinguish. However, the difference is that our solution is slightly affected, while the other two schemes, especially the distance-only one, are greatly affected at a low SINR threshold. This is because a low SINR threshold, namely a large effective communication range, increases the number of neighbors and the possibility of equidistant or constant velocity neighbor events. This is more severe in the small-size network. Besides, it is worth noting that changes in the SINR threshold have little impact on the performance of our dynamic weight scheme. Although the precision of the comparison schemes will be improved with the increase of the SINR threshold, they still fall behind our proposal with an average gap of 7.86\% and 8.95\%. 
\begin{figure*}
	\centering
	\includegraphics[width=0.9\linewidth]{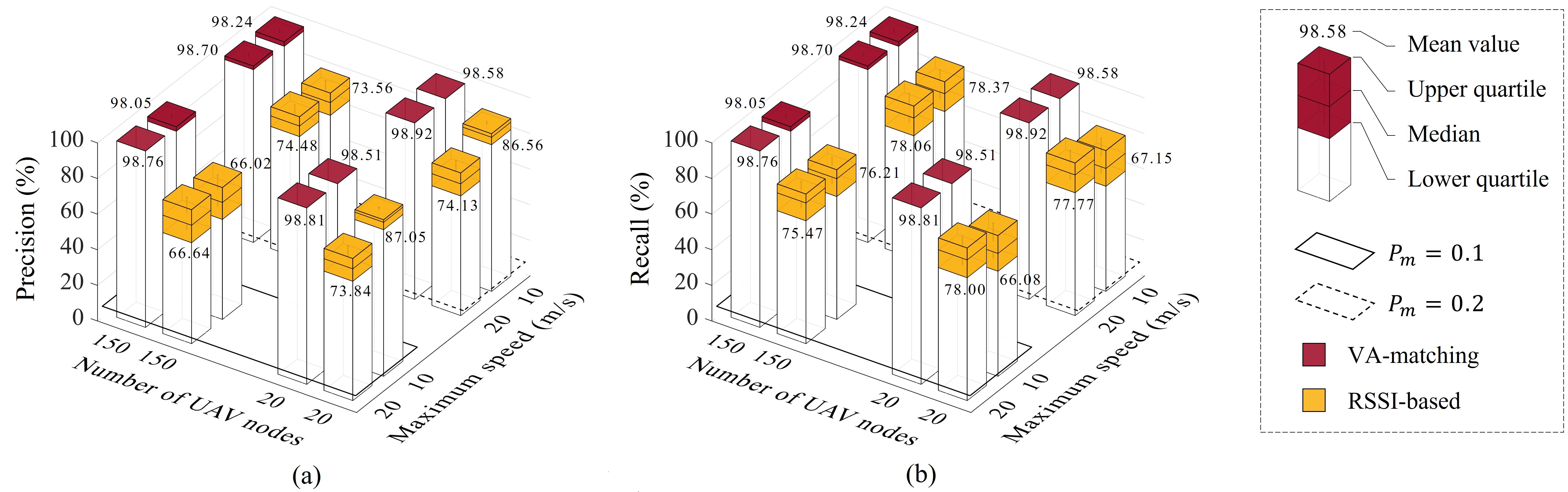}
	\caption{Sybil detection performance of VA-matching and RSSI-based method under different scenarios: (a) precision; (b) recall.}
	\label{Precision_Recall}
\end{figure*}

Finally, we compare the proposed VA-matching with the RSSI-based Sybil detection method. The principle of the RSSI-based method is: Messages sent by malicious nodes with different identities are constrained to have the same RSSI values at the receiver. If the RSSI of multiple neighbors are similar, they will be judged as Sybil nodes generated by the same malicious node. The RSSI ranging error model comes from the results of the real-world experiment. The detection recall is defined as R=TP/(TP+FN), where the false negative (FN) means the number of undetected Sybil nodes. 

As shown in \textbf{Fig. \ref{Precision_Recall}(a)}, both solutions have slightly better accuracy in a lower-density network for the following reasons. For one thing, matching errors hardly occur in VA-matching. For another, it is more difficult to have two neighbors with similar relative distances, i.e., similar RSSI, when using the RSSI-based method. In the high-density network, both solutions are insensitive to the maximum speed. This is because the number of neighbors and the topology change rate, which affects the matching accuracy and the possibility of RSSI similarity, are mainly affected by the network density rather than mobility in a dense network. In a sparse one, high-speed means worse precision for the RSSI-based method since the neighbor list change frequently. Consequently, new neighbors enter the communication range from all directions and undoubtedly increase the possibility of RSSI similarity. Our scheme is also insensitive to speed thanks to the high matching accuracy shown in \textbf{Fig. \ref{Matching_Rate}}. When the proportion of malicious nodes decreases from 0.2 to 0.1, the number of Sybil nodes has been reduced by half. As a result, among the detected results, the proportion of Sybil nodes that meet the RSSI threshold requirements decreases. Therefore, the precision of the RSSI-based solution scheme roughly decreases from 74\% to 66\%. Our solution is also not sensitive to this parameter. In conclusion, the detection accuracy of VA-matching is 23.29\% better than the RSSI-based one on average, and our solution has strong robustness.

Another important conclusion is that, unlike the conventional solutions, the precision and recall rates of our proposed VA-matching are equal rather than compromised. For the conventional solutions represented by the RSSI-based one, a higher recall is generally achieved at the cost of increasing the number of detected samples (e.g., increasing the threshold of RSSI difference to add more candidate nodes), which usually causes more legitimate nodes to be misjudged. It can be easily seen from the comparison results of \textbf{Fig. \ref{Precision_Recall}(a)} and \textbf{Fig. \ref{Precision_Recall}(b)} that in a network with 20 nodes, when the maximum speed is 20m/s, the RSSI-based method obtains a precision of 73.84\% and a recall of 78.00\%, while realizing a precision of 87.05\% and a recall of 66.08\% when the maximum speed is set to 10m/s. Recall that the difference in the number of neighbors in AD and VD is strictly equal to the real number of Sybil nodes. The VA-matching treats all the unmatched neighbors in AD as Sybil nodes, whose amount is equal to the number of real Sybil nodes. Consequently, the precision and recall are equal since they have the same denominators, which is of great significance in reality. Our solution can ensure that most of the Sybil nodes can be detected, and there will be almost no misjudgment of legitimate nodes.
\section{Conclusion}
In this paper, we have presented how to enable UAVs to accurately match what they ``see'' and ``hear''. The proposed VA-matching solution utilizes relative entropy to describe the similarity of characteristics and distinguishes UAVs by their popularity. Experiment results show that it yields better precision and recall when detecting Sybil attacks, and significantly outperforms the RSSI-based method. Given the page limit, we designate the detailed observation techniques in the dual domain and the real-world experiments of detecting Sybil attacks as our future work.

\section*{Acknowledgment}
This work was partly supported by Major Research Projects of the National Natural Science Foundation of China (92267202), the National Key Research and Development Project (2020YFA0711303), and the BUPT Excellent Ph.D. Students Foundation (CX2022208).

\vspace{12pt}

\end{document}